\documentclass{article}

\usepackage{PRIMEarxiv}

\usepackage[utf8]{inputenc} 
\usepackage[T1]{fontenc}    
\usepackage{hyperref}       
\usepackage{url}            
\usepackage{booktabs}       
\usepackage{amsfonts}       
\usepackage{nicefrac}       
\usepackage{microtype}      
\usepackage{lipsum}
\usepackage{fancyhdr}       
\usepackage{graphicx}       
\graphicspath{{media/}}     

\title{Artificial intelligence and downscaling global climate model future projections
\thanks{\textit{\underline{Citation}}: 
\textbf{Benestad. Artificial intelligence and downscaling global climate model future projections. arXiv:2601.00629.}} 
}

\author{
  Rasmus E. Benestad \\
  The Norwegian Meteorological Institute \\
  Oslo, Norway\\
  \texttt{\{Rasmus E. Benestad\} rasmus.benestad@met.no} \\
}

\begin{document}
\maketitle

\begin{abstract}
A critical review of artificial intelligence and deep machine learning (AI/ML) applied to downscaling of global climate model simulations provides some words of caution, based on past experiences and well-established principles. Recent papers tend to ignore more subtle successes with statistics and mathematical based downscaling, and there are examples of inappropriate evaluation strategies and incomplete accounts of the scientific progress when it comes to climate downscaling. An incomplete description state-of-the-art and a dogmatic approach to evaluation may give a deceiving impression that AI/ML is superior to more statistics and mathematics based methods.
\end{abstract}


\section{Introduction}

Artificial intelligence (AI) and deep machine learning (ML) have surged forward as a technology for weather forecasting and data analysis \cite{nipen_regional_2025,jin_advancing_2025,meraner_cloud_2020,ebiendele_deep_2025,pawar_esm_2024,rampal_high-resolution_2022,rampal_enhancing_2024,albu_nextnow_2022,fuentesfranco_paneuropean_2025,doury_regional_2021,leinonen_stochastic_2021,liang_disentangling_2025}, as well as many other uses, such as protein folding, chess playing, medical diagnostics, and the discovery of new materials \cite{jumper_highly_2021,silver_general_2018,esteva_dermatologist-level_2017,mckinney_international_2020,merchant_scaling_2023}. This rush towards AI/ML has been been accompanied by an impressive progress in large language models (LLMs), such as chatGPT, Gemini, Claude, and DeepMind, but these recent advances in AI/ML have also been hyped by a massive investments in Big Tech\footnote{BBC, Tech giants are spending big on AI in rush to dominate the boom, 30 October 2025, \url{https://bbc.com/news/articles/c5yp2y8rdpro}}.

Based on the impressive progress in AI/ML, one would naturally think that AI/ML also would be successful in making projections of future climate conditions. This is not a new idea, since AI/ML indeed was applied in climate research already in the 1990s and early 2000s. Past efforts on adopting AI/ML in downscaling global climate models (GCMs) include deep learning, artificial neural networks, and self-organising maps \cite{zorita_analog_1999,hewitson_self-organizing_2002,schoof_downscaling_2001}. Nevertheless, AI/ML didn't bring any breakthrough in climate research during the following two-to-three decades, and there may be several reasons for why it didn't become a preferred strategy within the downscaling community: (1) insufficient volumes with high-quality representative data, (2) the methods not being as developed then as the present ones, and (3) AI/ML is not really the best approach for this particular problem. 

In the early days of AI/ML, there was a concern that its methods constituted a "black box" approach for which we don't know why any specific results come out at the other end due to a lack of transparency \cite{hewitson_self-organizing_2002,wilby_guidelines_2004,benestad_downscaling_2010}.
There has since been developments in so-called "explainable AI" (XAI), for which an internal logic is visible and understandable in order to address this particular concern \cite{barredo_arrieta_explainable_2020}. 
The confidence in AI/ML as a method for studying climate change has also had a boost since then, particularly outside the climate downscaling community, and some of it may be due to successes within large language models, weather forecasting and other disciplines mentioned above. 
A landmark paper in 2017 is connected with more recent progress in using AI/ML in downscaling climate, involving super-resolution techniques to climate data \cite{vandal_deepsd_2017}. This paper discusses a transition from ESD to Deep Learning (DeepSD) and has set a standard baseline against which most modern methods are compared. 

On the other hand, concerns about an unfounded optimism in applying AI/ML uncritically to GCMs linger on and were expressed at the World Climate Research Programme (WCRP) COoRdinated climate Downscaling EXperiment (CORDEX) Science Advisory Team (SAT) open meeting in Hamburg November 18--19, 2025. 
In particular there are doubts about the use of AI/ML methods for downscaling climate change projections because AI/ML is not trained with data which represent the situation in a future climate \cite{gonzalez-abad_are_2025,hernanz_critical_2022,bano-medina_configuration_2020}. These concerns are further underscored by the discovery of instabilities in AI/ML methods for image recognition in medical sciences \cite{antun_instabilities_2020} and a faulty trait within AI/ML known as "hallucinations" \cite{raunak_curious_2021}.

Hence, in addition to the said concerns about {\em black box} and {\em hallucination}, the subtle but profound problem with AI/ML for studying climate change is the data itself. This concern can be paraphrased by the old saying "garbage in - garbage out" and is true for all models, be it GCMs, regional climate models (RCMs), empirical-statistical downscaling (ESD) or AI/ML. Furthermore, it won't go away with bias-adjustment or other types of post-processing. 

\section{Truly representative data}
\label{sec:representativedata}

While AI/ML is ideal for problem solving when it's trained on data which also are representative for the situation for which it is applied (despite being a black box that may hallucinate), it is more challenging when the training data no longer represent the context in which it is meant to provide predictions. In other words, AI/ML struggle if the situation or assumptions have changed between training and application, also referred to as "out-of-distribution performance" \cite{rampal_downscaling_2025}. 
This kind of non-stationarity is the core problem when it comes to climate change, and has long been recognised in ESD \cite{wilby_guidelines_2004,benestad_empirical-statistical_2008}. 

\begin{figure}
  \centering
  \includegraphics[angle=90,width=0.49\textwidth]{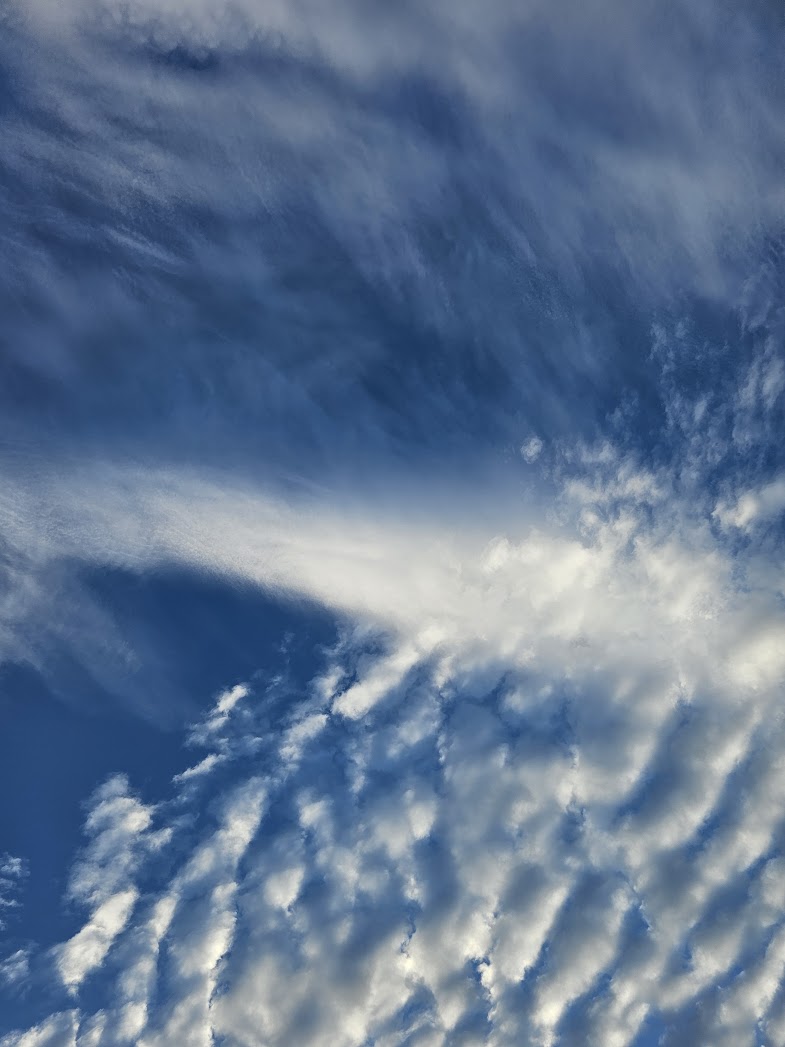}
  \includegraphics[width=0.49\textwidth]{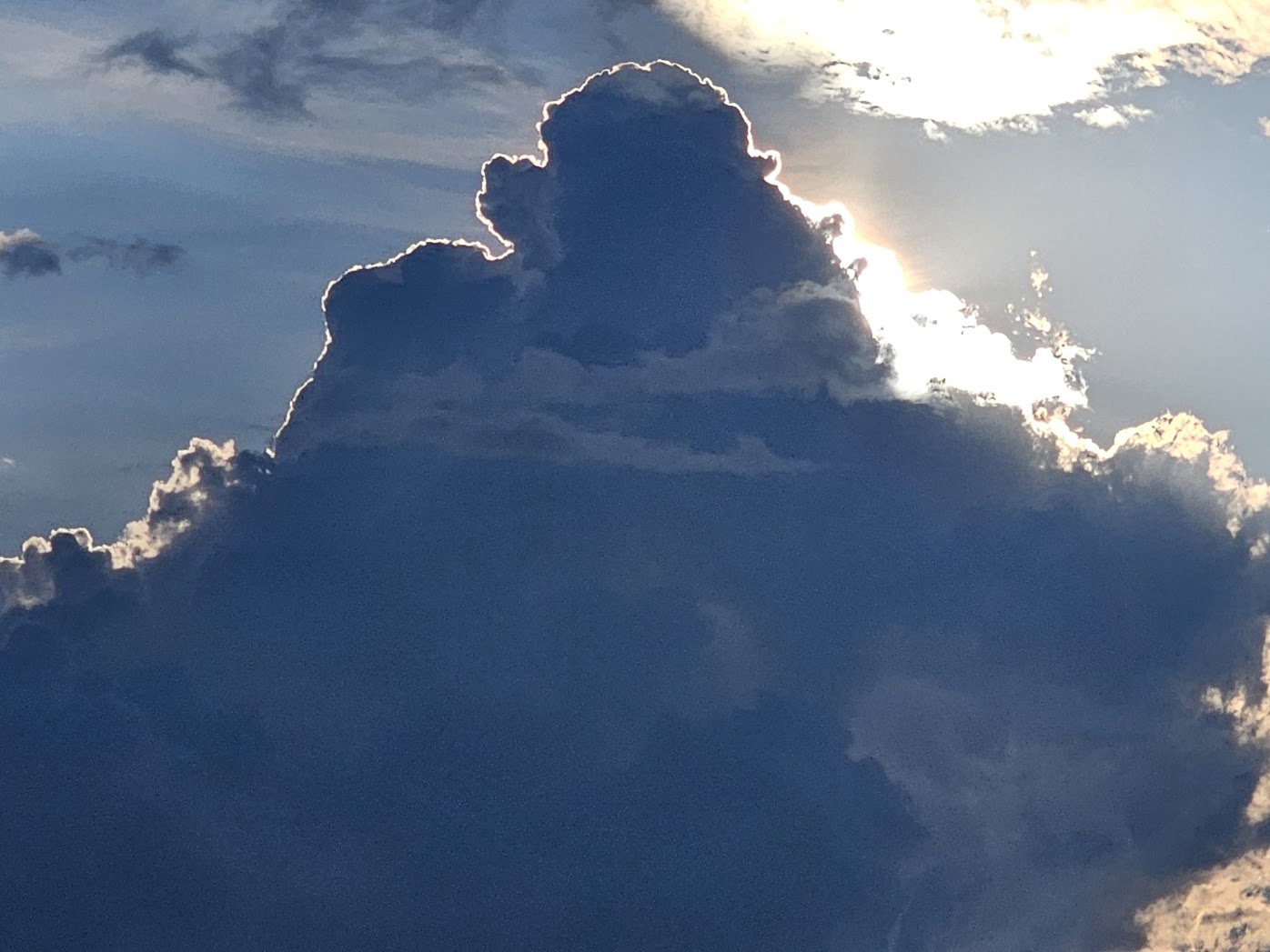}
  \caption{Clouds have many different scales, and different cloud types and structures illustrate different situations where links between large and small scales vary. They can also provide a metaphor for Earth's climate going through transitions from a known state to new unknown one with different internal structures. An AI/ML trained to predict small-scale features from the large scales in the clouds of left panel may not work well for those in the right hand panel. Often AI/ML involves cloud computing, pun intended.}
  \label{fig:fig1}
\end{figure}

The caveat concerning representativeness is illustrated by the cloud metaphor in Figure~\ref{fig:fig1} because clouds are a type of phenomenon which involves many different scales where the dependency between large and small scales in a cloud change as their structure and type vary. Clouds provide a good metaphor of climate, as they play a key role in the hydrological cycle, atmospheric dynamics, thermodynamics, radiative transfer, and ultimately rainfall patterns.
Another way to illustrate this is if we train AI/ML-based weather forecasts on winter rainfall and apply it for the summer season. 

Climate change and global warming are by definition the embodiment of changing conditions, so the internal relationships between various parts of the climate system in the past are expected to be different in the future.  Hence "climate change". Furthermore, GCMs are designed to simulate large-scale processes, phenomena and conditions within the climate system, how they interact, and how they change when subject to various forcings such as increased levels of atmospheric greenhouse gases. However, they are not constructed for providing the fine details needed when we want to know how global warming affects the local rainfall or temperature statistics. In other words, such models have a minimum skilful scale \cite{takayabu_reconsidering_2015}. The local climate is nevertheless depending on the large-scale ambient conditions and possibly teleconnections, and we can utilise the information about this dependency through a post-processing exercise known as {\em downscaling} \cite{von_storch_history_2019,gutowski_jr._wcrp_2016,benestad_downscaling_2016,gutman_comparison_2012,wilby_regional_2010,doblas-reyes_linking_2021,mearns_downscaling_2025,fowler_linking_2007}.

\section{Downscaling climate change}
\label{sec:downscaling climate}

The definition of downscaling in the scientific literature is often imprecise, and may sometimes be presented as an ambiguous step for merely increasing the resolution of the model output. 
It's useful to distinguish between (a) pure interpolation to finer grid-spacings, (b) so-called bias-adjustment, and (c) downscaling. An interpolation is just that - an interpolation, and bias-adjustment is merely a step that adapts the model output so that it has similar statistics as corresponding "truth" \cite{gudmundsson_technical_2012,maraun_bias_2013}. Downscaling, on the other hand, involves making use of the information describing the link between the large-scale aspects, which GCMs skilfully reproduce, and small scales (hence its name) \cite[appendix]{benestad_downscaling_2025}. 
Climate models have a so-called minimum skilful scale \cite{takayabu_reconsidering_2015}, and statements such as "this mapping, which must transform a single grid point to multiple points is an ill-posed problem, one with many possible solutions" \cite{vandal_deepsd_2017} suggest that the minimum skilful scale is often not considered in papers on downscaling.
This definition also indicates why out-of-distribution performance is a challenge for downscaling, but not necessarily for interpolation or bias-adjustment. Downscaling is a broad term which may involve RCMs, ESD, dynamical methods, as well as AI/ML that take large-scale information as input to infer a local response.

Both ESD and AI/ML methods can be trained on data produced by models, for instance high-resolution output from RCMs. When such simulations are used to test ESD and AI/ML methods it is called {\em perfect model experiments}. It is also possible to use methods trained on RCM results for making more detailed projections for the future, and this is referred to as {\em emulation} \cite{rampal_downscaling_2025} or hybrid ESD-RCM downscaling \cite{erlandsen_hybrid_2020}.
One caveat, however, is that RCMs only provide an approximate representation of reality, and different RCMs can give different accounts concerning climatic condition, even when subject to the same large-scale boundary conditions \cite{nikulin_precipitation_2012,kotlarski_regional_2014}. 
Furthermore, comparisons between observations and simulations often expose systematic biases \cite{orskaug_evaluation_2011,benestad_complex_2007}, and there will always be a need for observations in bias-adjustment \cite{gudmundsson_technical_2012}. In addition, the RCMs make so-called hydro-static approximations where convection is not explicitly simulated, but recently so-called convection-permitting models (CPMs) have been introduced which give a different picture on the summer time rainfall \cite{kendon_realism_2012}. 
The questions are still open whether these imperfections affect our ability to test ESD and AI/ML, or how they limit the representation of emulators.

There are also three different strategies used in ESD: (1) perfect prognosis (PP), (2) model output statistics (MOS), and (3) hybrid PP-MOS. The most widely used strategy studying climate change has traditionally been PP \cite{maraun_value:_2015}, whereas MOS is more suitable for seasonal or decadal forecasting \cite{benestad_using_2019}. 
However, the transferability/portability for a method trained on observations to a model universe is not trivial, and a hybrid PP-MOS scheme involving {\em common empirical orthogonal functions} (EOFs) has therefore been proposed and is based on a joint observation-model data framework \cite{benestad_interaction_2001}. Such common EOFs are also suitable for capturing the large-scale covariance which GCMs are able to reproduce with significant skill \cite{benestad_various_2023}. The challenge associated with transferability/portability also implies that the training data may not be entirely representative for the use with model output, especially if the model is not stable or robust.

\section{Proper testing and benchmarking}
\label{sec:testing}
 
It's important to test whether the AI/ML methods provide relevant and robust results as well as their capabilities, and there has been a tendency to use ESD as a benchmark in downscaling studies. Most such efforts, however, have traditionally only involved ESD efforts that downscale weather rather than climate \cite{rampal_enhancing_2024}. For instance, a study with deepSD used the years 1980--2005 for training (9496 days) while 2006--2014 (3287 days) were used for validation \cite{vandal_deepsd_2017}. The difference between {\em downscaling weather} and {\em downscaling climate} may seem subtle at first sight, but is more profound when taking into account how they represent different solutions. 
The former is similar to weather forecasting, where the model makes step-by-step prognoses for a sequence of outcomes and where each outcome is represented by a data point. 

For climate change studies, on the other hand, the purpose is not to predict the exact weather evolution, but how the weather statistics change, and thus the traditional weather approach is no longer optimal for this objective. A change in strategy is described by \cite{benestad_norwegian_2025} where the objective is to predict how the shape of statistical functions describing probabilities change, as opposed to changes to individual data points. The motivation behind this is that statistical properties, such as parameters of curves and probabilities, are generally more predictable than individual outcomes. For instance, it's easier to estimate the probability of non-zero rainfall for a specific location than it is to predict if it's going to rain on a particular day. There are some examples involving the downscaling of wind speed probability distributions, probabilities of heavy 24-hr precipitation and associated return periods, in addition to heatwave statistics \cite{pryor_empirical_2005,pryor_climate_2005,pryor_winds_2006,benestad_downscaling_2025,benestad_downscaling_2018}.
 
Another tendency has been to test downscaling with AI/ML algorithms against ESD based on present weather, but not their ability to simulate an out-of-sample future climate when applied to GCM results \cite{rampal_enhancing_2024,wouters_exsodos_2025,kashinath_physics-informed_2021}. In such a setting, it is expected that AI/ML will outperform ESD when it comes to simulating present conditions because it has many more parameters that can be tuned, but such tests will not provide any information about how reliable the results will be for a future climate. Such exercises may be motivated by the protocol recommended by the European COST-Value project \cite{gonzalez-abad_are_2025,bano-medina_configuration_2020} which merely tests the methods in an ideal setting for the past with predictors based on reanalyses \cite{maraun_value:_2015}. 

It's also common to involve a mix of predictor variables to predict individual points of a predictand variable, but a climate change often introduces non-stationarities within the internal structure of the climate system, and hence between variables and levels in a global climate model \cite{parding_statistical_2019}. One question is which part of the predictor carries more weight if they diverge. However, it is also possible to constrain the predictors to a single variable to estimate the parameters of mathematical functions representing the statistical properties of variables, and such an approach puts more weight on mathematical theory as well as more constraints on data fit \cite{benestad_downscaling_2025}. 

\section{Data and mathematics}
\label{sec:testing}

Data may be scarce, have errors or may not represent exactly the conditions in which we are interested, and therefore represents a limitation to AI/ML and other data learning strategies. Mathematics, on the other hand, are absolute and true everywhere and to all times, and the information and constraints by the mathematical framework can compensate for lacking and imperfect data. ESD that is based on mathematical rules and statistical theory can therefore provide more robust information when there are imperfect or scarce data. Moreover, this theoretical basis has been a motivation for the concept "ESD", empirical-statistical downscaling, rather than just empirical downscaling or just statistical downscaling. 

\section{Good use of AI/ML for climate downscaling}
\label{sec:good-use}

There is no doubt that AI/ML is impressive and has many good uses, although past experience and theoretical concerns call for caution when used for downscaling future climate conditions. One solution is to use AI/ML for emulating RCM simulations, and especially for studying internal variability \cite{rampal_downscaling_2025}. Deep-learning downscaling algorithms are able to reproduce simulated future climate change responses accurately through emulation, both for mean as well as extreme precipitation, but this is only true when they have been trained on climate simulations that include future periods \cite{rampal_extrapolation_2024}.
Such emulations have been applied to Single Model Initial Condition Large Ensembles (SMILEs) with ensemble sizes with up to $\sim 40$ members in order to eliminate issues concerning transferability/portability, as it's not straight forward to apply an emulator trained on a specific GCM to another GCM with different quirks \cite{rampal_downscaling_2025}.

One point often missed by AI/ML studies is that properly designed ESD efficiently can downscale more than 1000 runs in a few hours on an ordinary server or even a laptop \cite{schuler_svalbards_2025} which is more efficient than AI/ML ($\sim 48$ hours of training on GPUs and more powerful computers) \cite{rampal_reliable_2025}. Furthermore, while AI/ML may involve an order of million trainable parameters for precipitation, ESD can be manage with two for daily precipitation statistics (wet-day frequency $f_w$ and wet-day mean precipitation $\mu$), each represented by typical the seven leading modes of common EOFs \cite{benestad_downscaling_2025}. Hence, ESD may also give more robust results with carefully designed methods and strategy.
Furthermore, downscaling the shape of statistical curves (their parameters, e.g. for probability density functions or plain probabilities) gives information about extremes (the tail of the distributions) and is not subject to the same caveat as downscaling individual data points (outcomes) that is referred to as "regression-to-the-mean"\cite{rampal_enhancing_2024}.

In addition to emulation, AI/ML may also provide promising techniques for extracting features and patterns in huge data volumes \cite{benestad_new_2017}.
Another potential applications may include simulating storm tracks, the meandering of the polar jet stream, seasonal forecasting at mid-to-high latitudes, and the shape of statistical curves. There has been some work using quantile regression neural network in downscaling precipitation \cite{cannon_quantile_2011}, and AI/ML algorithms such as cGANs may be used in stochastic weather generators, a possibility that apparently has been both under-utilized and under-evaluated in climate science \cite{rampal_reliable_2025}.

\section{Discussion}
\label{sec:discussion}

The recent success with AI/ML is partly brought on by vast data volumes and increased computational capabilities, in addition to new methods, and can be described a result from 'brute force' approach to data that involves tuning millions of parameters and requires an infrastructure of a different kind than ESD, which can be carried out on a laptop or a normal server \cite{benestad_norwegian_2025}. 
ESD involves "human intelligence" to a greater extent because it's based on learning from other relevant studies which may not necessarily be directly associated and easy to miss with AI/ML. One example is the patterns of daily precipitation from around the world \cite{benestad_spatially_2012,benestad_specification_2012}. In addition, basing ESD on our understanding of physics as well as statistics, gives it an advantage that is not as easy to exploit with AI/ML \cite{kashinath_physics-informed_2021}. 

Most papers on ESD and downscaling with AI/ML involve the 'downscaling weather' strategy which is susceptible to the said 'regress-to-the-mean' type behaviour and an inability to properly capture extremes \cite{rampal_reliable_2025}.
Again, this caveat is not a problem for strategies which aim to predict the shape of the curves representing statistical distributions, because the extremes are represented by the tails of these curves. It is also expected that ESD and AI/ML don't perform as skilfully with out-of-sample data as they do during their training, which is especially a concern when they are applied to a new type of reality such a changed climate. 
Therefore it's important to test the models' ability in terms of projecting a future climate change when applied to different GCMs and not only whether they reproduce the statistics of historical data \cite{wouters_exsodos_2025}. 

There have been some studies which assess whether machine AI/ML can be used to extrapolate results for a future climate \cite{rampal_downscaling_2025}, and one suggests that some methods underestimate climate change signals for extreme precipitation despite providing impressive visual 'realism' \cite{ward-leikis_intercomparison_2025}. Another study suggests that several algorithms employed in AI/ML do not simulate well a future climate which is different to the one on which they were trained and tried improve their capabilities by imposing physical constraints \cite{kashinath_physics-informed_2021}.
In this context it is useful to keep in mind that out-of-sample is not exactly the same as the question of representativeness, as the latter also applies to models trained on the historical climate which may not be valid for a future climate that is different.
The out-of-sample problem is not similar for parameters of distributions as for individual data points, since models designed to predict such statistical parameters assume changed statistical properties for temperature or precipitation in their design.

When ESD is used to benchmarked AI/ML it is important to compare with relevant, suitable and the best methods, and not pick a random or sub-optimal performer, which often seems to be the case. Hence, incomplete accounts on the state-of-the-art within downscaling and biased referencing to past work may lead to inaccurate conclusions \cite{benestad_learning_2015}.
In other words, it's important to avoid using "straw man" methods in the benchmarking. 
A proper protocol must be in place to avoid maladaptation \cite{ipcc_climate_2021}, which includes representative references to the scientific literature and past experiences, as climate change adaptation involves expensive investments. It will take decades before one knows whether the projections turns out to realistic, which is different to the case where predictions can be assessed against the actual outcome, such as in weather forecasts and seasonal forecasts where the actual outcomes are available on a frequent and regular basis for evaluation and further learning.

An often unspoken concern with AI/ML is that it also may invite to a more dogmatic attitude, taking the results by their face value, as opposed to a more scientific attitude where transparency and replication are important as well as the understanding of the system and processes behind the results.

\section{Conclusion}
\label{sec:conclusion}

While AI/ML has had impressive success in various fields, it is not guarantee that it will be as successful when it comes to downscaling simulations of a future by global climate models. Past studies have exposed limitations and there are mathematics and statistic-based methods which may give more accurate results. Often relevant experience from decades of work in empirical-statistical downscaling is not accounted for in studies of AI/ML capabilities, and it's important to use the best alternatives for benchmarking because it's no challenge to find a poor straw-man method that anyone can beat. A real test is to beat the best alternatives, and when the method is used to compute climate change, it's important to apply appropriate tests which evaluate the skill for intended use. 

\section*{Acknowledgments}
This paper was not supported by any grants, and was written on a Linux laptop with Overleaf, and Zotero was used for handling references. The manuscript was loaded up to Gemini to check whether it had missed any seminal paper relevant in this topic, and Gemini and Ecosia were used to search for previous work relevant for this discussion and their DOI. However, in some cases Gemini hallucinated and claimed that a paper discussed questions which it really didn't.

\bibliographystyle{unsrt}  
\bibliography{references}

\end{document}